\documentclass[12pt]{iopart}

\usepackage{textcomp}
\usepackage{graphicx}
\renewcommand{\figurename}{\textbf{Figure}}
\usepackage{fixltx2e}

\begin{document}
{
\title[]{Statistical measurements of quantum emitters coupled to Anderson-localized modes in disordered photonic-crystal waveguides.}
\author{Alisa Javadi\footnote{Authors to whom any correspondence should be addressed.}, Pedro D. Garc\'{i}a, Luca Sapienza\footnote{Current address: School of Engineering and Physical Sciences, Heriot-Watt University, Edinburgh, United Kingdom.}, Henri Thyrrestrup\footnote{Current address: MESA+ Institute for Nanotechnology, University of Twente, Enschede, Netherlands.} and Peter Lodahl\setcounter{footnote}{1}\footnotemark}
\address{$^1$Niels Bohr Institute,\ University of Copenhagen,\ Blegdamsvej 17,\ DK-2100 Copenhagen,\ Denmark\\}
\ead{Javadi@nbi.ku.dk, Lodahl@nbi.ku.dk}
\date{\today}

\begin{abstract}

Optical nanostructures have proven to be meritorious for tailoring the emission properties of quantum emitters. However, unavoidable fabrication imperfections may represent a nuisance. Quite remarkably, disorder offers new opportunities since light can be efficiently confined by random multiple scattering leading to Anderson localization. Here we investigate the effect of such disorder-induced cavities on the emission dynamics of single quantum dots embedded in disordered photonic-crystal waveguides. We present time-resolved measurements of both the total emission from Anderson-localized cavities and from single emitters that are coupled to the cavities. We observe both strongly inhibited and enhanced decay rates relative to the rate of spontaneous emission in a homogeneous medium. From a statistical analysis, we report an average Purcell factor of 2 in without any control on the quantum dot - cavity detuning. By spectrally tuning individual quantum dots into resonance with Anderson-localized modes, a maximum Purcell factor of 23.8 is recorded, which lies at the onset of the strong coupling regime. The presented data quantify the potential of naturally occurring Anderson-localized cavities for controlling and enhancing the light-matter interaction strength, which is of relevance not only for cavity quantum-electrodynamics experiments but potentially also for efficient energy harvesting and controllable random lasing.
\end{abstract}


\section{Introduction}

\begin{figure}[h!]
  \includegraphics[width=155mm, height=105mm]{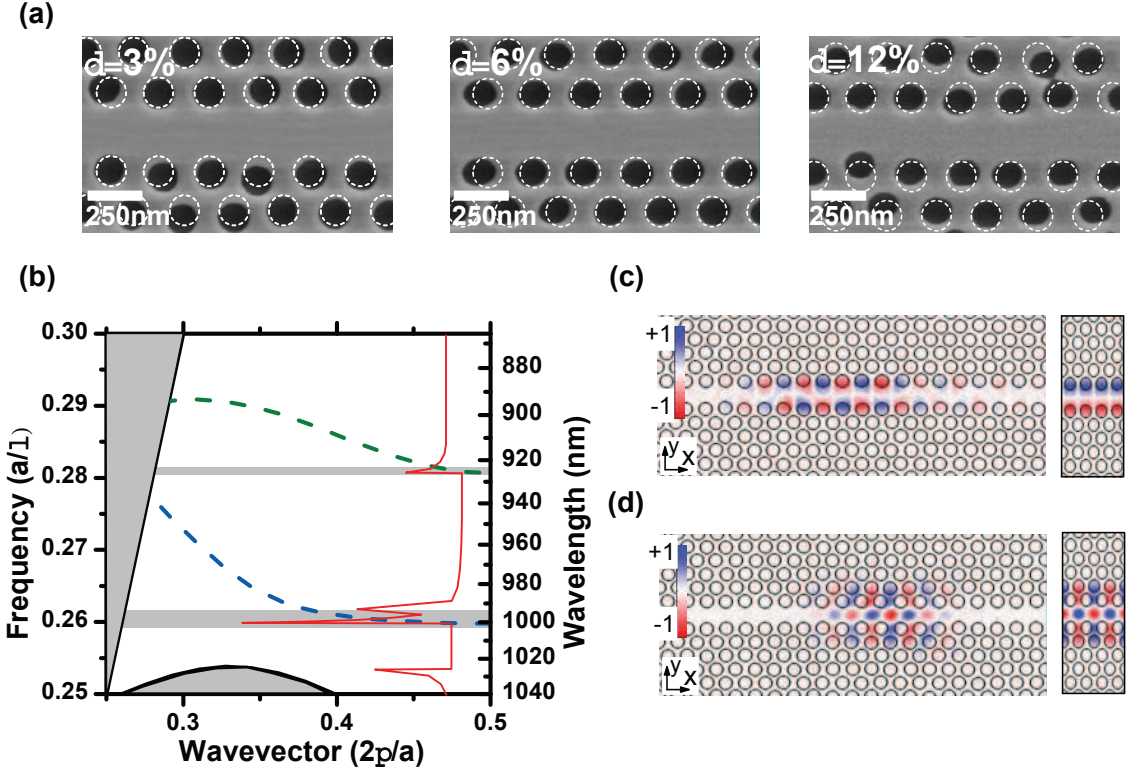}
      \caption{ \label{1} \textbf{Anderson localization in photonic-crystal waveguides.} (a) Scanning-electron micrographs of photonic-crystal waveguides with different amounts of intentional disorder in the hole position. Dashed circles indicate the positions of the holes in a perfectly periodic structure. (b) Dispersion relation for an ideal photonic crystal waveguide showing the fundamental (blue) and high-energy (green) guided modes. The shadowed area near the cutoff of the guided modes indicate the spectral range where Anderson-localized modes appear. The red curve is a sketch of local density of optical states of a disordered structure. (c) and (d) Two-dimensional finite-element calculation of the $E_{y}$ component of the electric field for an Anderson localized mode along a perturbed PhCW with $\sigma=1\%$ engineered disorder in the hole positions (left) and along a perfect PhCW (right) corresponding to the high-energy ($\lambda=850$ nm) (c) and fundamental ($\lambda=930$ nm) (d) guided modes shown in (b).}
    \end{figure}

The local environment of a quantum emitter determines its spectral and spatial emission properties. Within the dipole approximation, the emission dynamics of an emitter is directly determined by the local density of optical states (LDOS) \cite{novotny}, which accounts for the density of vacuum fluctuation at the position of the emitter. By tailoring the LDOS it is possible to enhance or inhibit the emitter spontaneous-emission rate through the so-called Purcell effect \cite{purecell}. For a sufficiently large coupling strength a single emitter and a single electromagnetic mode can become entangled, which is referred to as strong coupling. Among the different approaches for tailoring the LDOS by nanoengineering, photonic crystals (PhC) have so far been the most successful and controlled spontaneous emission \cite{PFinPHC,peter} with modifications of the emission rate approaching two orders of magnitude \cite{qin} have been demonstrated together with strong coupling between a single quantum dot and a photon in a photonic-crystal cavity \cite{scinphc}. PhCs are dielectric structures where a periodic variation of the refractive index leads to the formation of a frequency range, the photonic bandgap, where electromagnetic wave propagation is strongly suppressed \cite{Joannopoulos}. In frequency ranges outside the bandgaps, light propagation is described by periodic Bloch modes. One possible implementation of a two dimensional PhC is obtained by etching a triangular lattice of holes in a membrane of a high refractive index material. In such a platform, a photonic crystal waveguide (PhCW) is formed by leaving out a row of holes, see Fig. \ref{1}(a). In that case, the light is tightly confined and effectively guided along the missing row of holes due to the presence of an in-plane bandgap in the PhC and by total internal reflection within the membrane. Two different (longitudinal) waveguide modes are found in the bandgap, cf. the dispersion diagram of Fig. \ref{1}(b) and the mode profiles in  Fig. \ref{1}(c) and (d). At the edges of a PhC bandgap and close to the the cut-off of the waveguide modes, the LDOS is strongly enhanced and in the ideal case of a perfect crystal it would diverge, which is the so-called van-Hove singularity implying that the group velocity of the waveguide mode vanishes and a standing wave is formed. In reality imperfections and loss will smoothen the singularity, however a strongly enhanced LDOS prevails near the cutoff of the waveguide mode \cite{Henri_apl}.  This ability to enhance the LDOS makes PhCs and PhCWs very promising for slowing down light \cite{baba}, optomechanical experiments \cite{painter} and efficient single-photon sources \cite{toke} for quantum-information applications.

\begin{figure}[h!]
  \includegraphics[width=157mm, height=160mm]{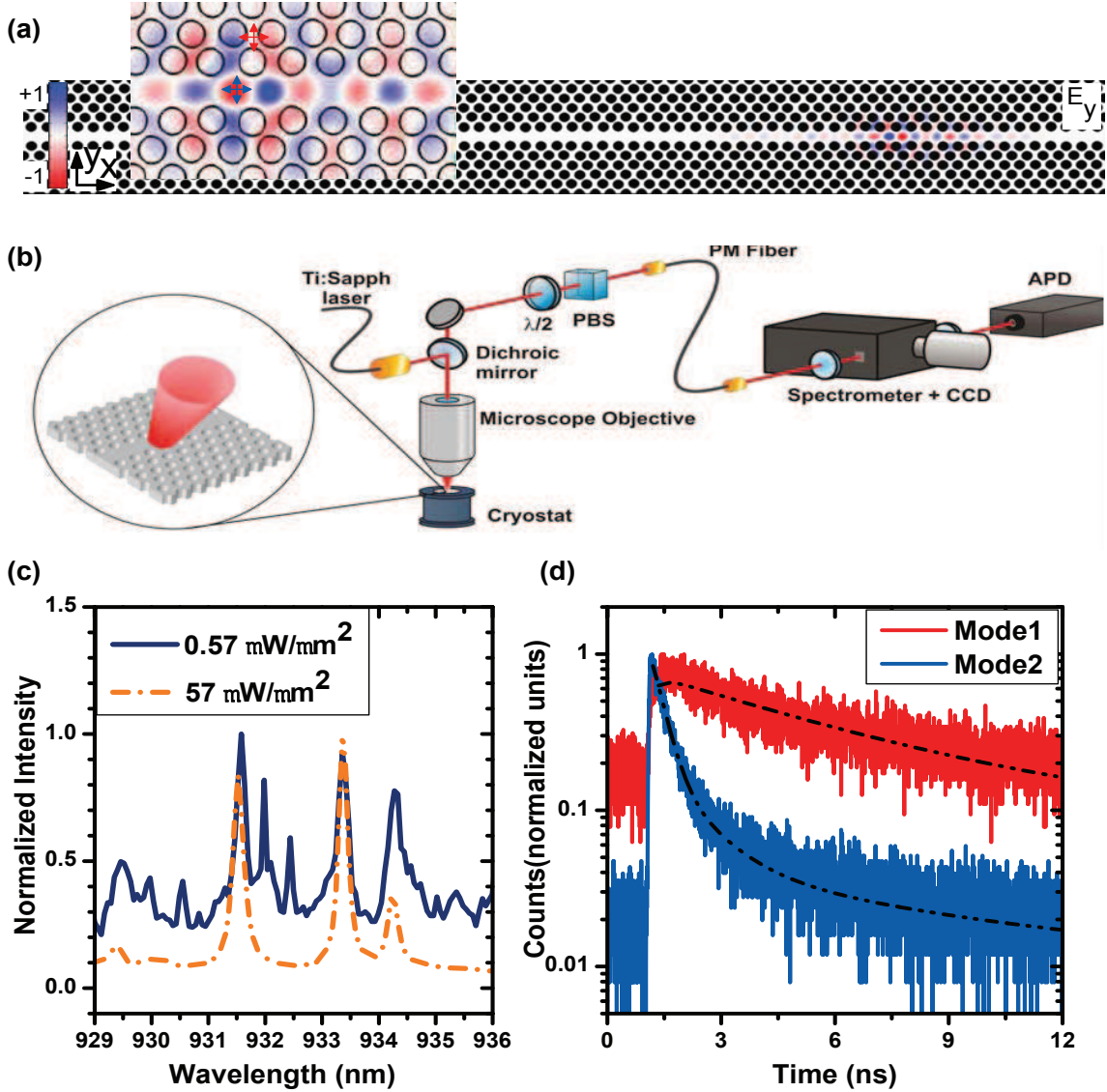}
      \caption{ \label{2} \textbf{Samples and experimental method.} (a) Two-dimensional finite-element calculation of the $E_y$ component of an Anderson-localized mode along a PhCW perturbed with $\sigma=1\%$ introduced disorder. The inset shows two dipoles corresponding to two quantum dots placed at a node (red) and an anti-node (blue) of the cavity, thus experiencing a very different local density of optical states. (b) Sketch of the experimental setup. See main text for detailed explanations. (c) Emission spectra of the sample under high excitation power (dashed curve), showing the Anderson localized modes and under low excitation power ( solid curve) showing Anderson localized modes and quantum dots. (d) Examples of time-resolved photoluminescence decay curves of different Anderson-localized cavities fitted with multi exponentials (dashed lines). The very pronounced differences in the decay times are attributed to the different spatial and spectral positioning of the dominant emitter feeding the cavities.}
    \end{figure}

The presence of disorder in a PhCW, ultimately  due to the limited precision of fabrication processes, breaks the discrete translational symmetry. Disorder induces random multiple scattering of light that removes van Hove singularities and may create randomly localized modes by Anderson localization \cite{SJohn,PAnderson} that approximately inherit the polarization properties of the fully periodic waveguide as seen in Fig. 1c and 1d. Due to their random nature, a statistical analysis is required to extract the spectral and spatial properties of Anderson-localized modes. They appear around and below the cut-off frequency of the waveguide mode or at the bandedge forming a so-called Lifshitz tail \cite{savona,huisman}, as marked in Fig. \ref{1}(b) by the shadowed regions for the waveguide modes. After ensemble averaging over all configurations of disorder, the electric field from an embedded emitter will decay exponentially in space with a characteristic length called the localization length ($\xi$) \cite{sheng}. A finite-element calculation of the E\textsubscript{y} components of two different Anderson-localized modes is shown in Fig. \ref{1}(c and d) that is compared to the Bloch modes of the ideal periodic structure. Remarkably, such random cavities in a PhCW have been found to have quality (Q) factors and mode volumes that are comparable to state-of-the-art engineered cavities  \cite{highq,highq2,akahane} with the obvious benefit of having less stringent requirements on sample fabrication precision. Disorder-induced cavities have attracted significant attention and have been proposed for light harvesting application \cite{wirsema},  used in cavity quantum electrodynamic (QED) experiments  {\cite{luca_science}} and for random lasing \cite{jin,tureci}.

The probability of entanglement between a single quantum dot and a photon in an Anderson-localized mode in a disordered PhCW has been investigated in Ref. \cite{henri_prl}, where a probability of $1\%$ was found for parameters corresponding to present experiments. In addition, from the statistical distribution of Purcell factors the probability to observe highly enhanced decay rates was assessed. In this paper, we present statistical measurements of the decay dynamics of both single quantum dots tuned to resonance with Anderson-localized modes and for the case where no control over the quantum dot - cavity detuning is implemented. The data set provide two alternative ways of extracting experimentally the Purcell factor statistics that was theoretically analyzed in \cite{henri_prl}. In the first presented data set we measure the decay dynamics of the Anderson-localized modes and extract the fastest rate of the multi-exponential decay curves. This procedure records the rate of the quantum dot that is best coupled to this particular Anderson-localized mode while the detuning between the emitter and the cavity is not optimized. For the second data set we monitor the decay rate of a single exciton line while temperature tuning it into resonance with an Anderson localized mode. This uncovers the full potential of the Anderson-localized modes but at the expense that the measurements are very time consuming thereby limiting the amount of statistical data that can be collected. From our measurements, we evaluate the light-matter interaction strength and compare the experimental data to the theoretically predicted distributions illustrating that the best cavities are at the onset of the strong-coupling regime.

\section{Experimental Methods}

The samples studied are 150 nm thick GaAs membranes with an embedded layer of self-assembled InAs quantum dots in the center with a density of 80 $\mu${m}\textsuperscript{-2} that emit light in the 890 nm - 1000 nm wavelength range. A set of 100 $\mu${m} long PhCWs with a triangular lattice of holes and varying lattice constant ($a$) and hole radius ($r$) are etched on the membrane. All waveguides are at least 10 times longer than the measured localization length \cite{njp_smolka}, meaning that Anderson-localized modes are formed near the cut-off of the waveguide mode. In addition to intrinsic fabrication imperfections in shape, size, and position of the holes, additional engineered disorder is introduced in the sample by randomly varying the position of the three rows of holes on each side of the waveguide according to a normal distribution with a mean value of zero and a variance of $\sigma \times a$ where $\sigma$ is varied from $0\%$ to $12\%$ (cf. Fig. \ref{1}(a)). Figure \ref{2}(a) schematically represents two quantum dots at two different positions showing the two potential dipole orientations with respect to the E\textsubscript{y} field component of an Anderson-localized mode in a PhCW. For carrying out the optical measurements, the sample is placed in a liquid Helium flow cryostat and cooled down to 10 K, see Fig. \ref{2}(b). A pulsed Ti:Sapphire laser emitting at 800 nm is focused on the sample through a microscope objective with NA = 0.55 from the top to a spot size of about 1.4 $\mu$m\textsuperscript{2} and the emission from the quantum dots is collected through the same microscope objective. The cryostat is mounted on translational stages to control the excitation and collection positioning with an accuracy of 100 nm. The emission is polarization filtered with a halfwaveplate and a polarizing beamsplitter, coupled to a polarization maintaining single mode fiber for spatial filtering, and sent to a monochromator with spectral resolution of 50 pm. The filtered light is finally detected with a CCD for spectral measurements or with an avalanche photo diode (APD) in time-resolved measurements.

 Time-resolved measurements are performed using two different approaches. In the first one, a set of waveguides with lattice constant a = 240 nm, hole radius r = 74 nm, and different disorder degrees (0\%, 1\% and 2\%) are investigated, where the cut-off of the fundamental guided mode is at 930 nm. For high pump powers (57 $\mu$W/$\mu$m\textsuperscript{2}), the spectral properties of the Anderson-localized modes can be determined. The excitation power is then reduced to 0.57 $\mu$W/$\mu$m\textsuperscript{2}, cf. Fig. \ref{2}(c), which is close to the saturation power of a single quantum dot and time-resolved measurements are performed on the cavity emission spectrum. In such time-resolved measurements on the cavity peak, emission is recorded from all the quantum dots that are coupled to the cavity mode implying that the decay curves are generally multi-exponential and we concentrate here on the fastest component of the decay curves corresponding to emission from the quantum dot that couples best to the cavity.{ The measured decay curves (see Fig. \ref{2}(d) for representative examples) are fitted satisfactorily well with either single exponential, bi-exponential or triple-exponential models after convolution with the 66 ps wide instrument response function of the setup.}  The same procedure is repeated for all of the observed Anderson-localized modes in the samples and very large variations are observed between different cavities. This procedure enables us to acquire a large data set for the statistical purposes, which provides a reliable lower bound on the actual Purcell enhancement that can be obtained in the system, since the detuning between the quantum dots and cavity modes is not controlled. In the second approach the Purcell factor is probed directly by time-resolved photoluminescence experiments on a single quantum dot emission line after tuning it into resonance with an Anderson-localized cavity by varying the temperature from 10 K to 30 K \cite{kiraz}. The fast decay rate originates from recombination of the bright exciton of the resonant quantum dot. The Purcell factor is extracted by relating the measured decay rates to the average decay rate of 1.1 ns\textsuperscript{-1} obtained {from quantum dots} in homogeneous environment. The optimum Purcell factor for a quantum dot perfectly matched spatially and spectrally to a cavity is given by $F_P=3Q(\lambda/n)^3/4\pi^2V$, where n = 3.44 is the refractive index of the membrane and $Q$ and $V$ are the quality factor and mode volume of the cavity. From this relation a conservative upper bound on the mode volume of the Anderson-localized cavity can be extracted.

\begin{figure}[ht!]
  \includegraphics[width=157mm, height=140mm]{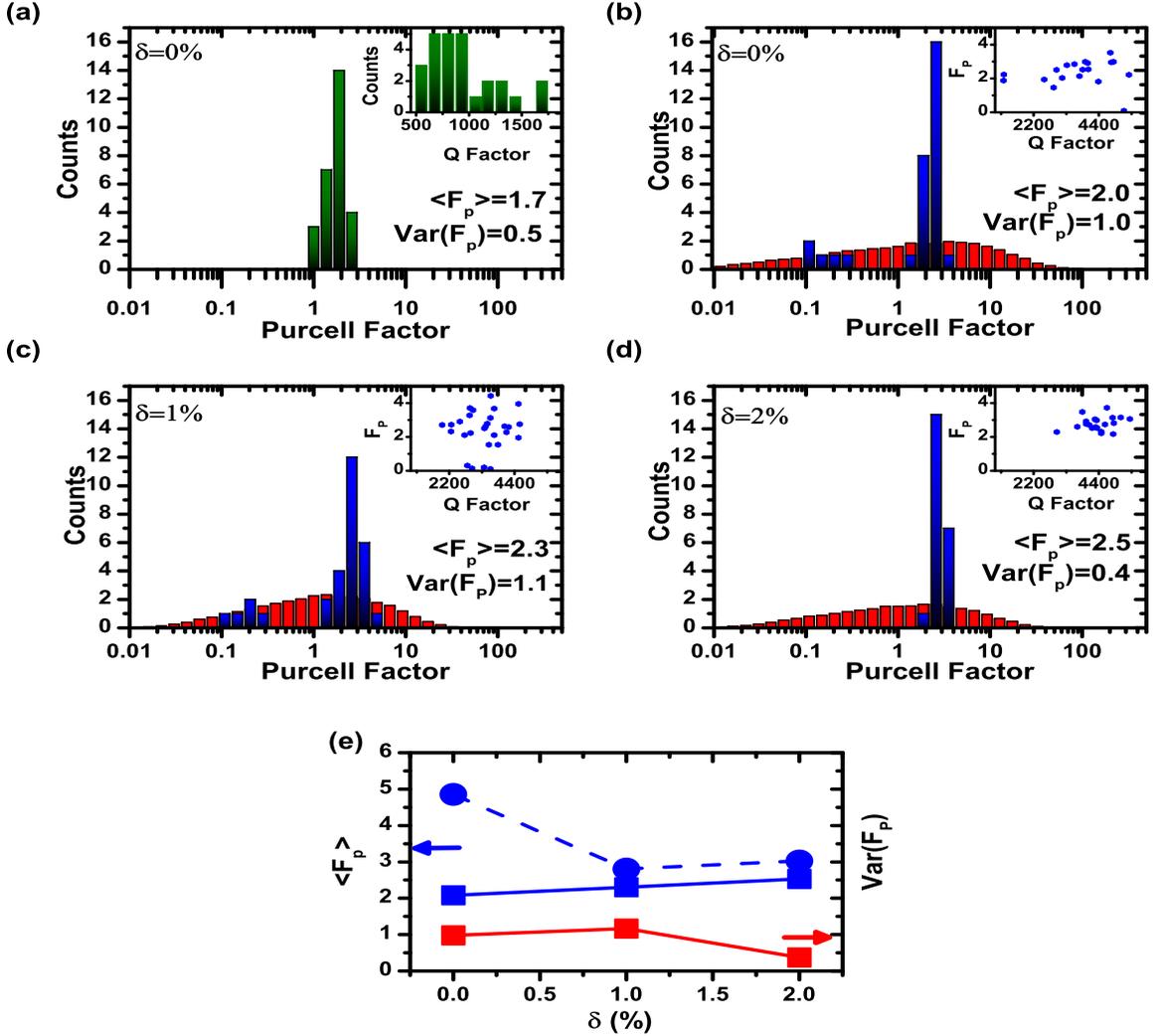}
      \caption{ \label{3} \textbf{Statistics of decay rates measured on the Anderson-localized cavities} (a) Histogram of the measured decay rates from Anderson-localized modes appearing along an unperturbed PhCW ($\sigma=0\%$) with a = 260 nm and r = 78 nm for the  high-energy guided mode. The inset shows the histogram of the cavity Q-factors. (b),(c) and (d) Histograms displaying the Purcell factor measured on Anderson-localized modes for the fundamental guided mode of a PhCW with a = 240 nm and  r = 74 nm and for $\sigma=0\%$, $\sigma=1\%$  $\sigma=2\%$, respectively. The blue histograms correspond to the experimental measurements while the { red} histograms are the corresponding distributions obtained from the theory in \cite{henri_prl}. The theory and experiment histograms are scaled in order to have the same {area}. The insets show the measured decay rates vs. the corresponding cavity Q-factor. The lack of correlation between Q factor and decay rates is attributed to the uncontrolled spatial and spectral matching of the dominant emitter to the cavity. (e) Mean and variance of the measured Purcell factors vs. degree of disorder degree (squares) and the corresponding calculated mean value (circles) for the data of (b)-(d). }
    \end{figure}

\section{Time-resolved measurements on Anderson-localized cavities}

In the following we present the experimental data of the spontaneous emission dynamics recorded when collecting light from the Anderson-localized cavities. We study the two different waveguide branches and for different degrees of disorder. Figure \ref{3} shows the statistics of the measured Purcell factor. The histogram in Fig. \ref{3}(a) shows the case of the secondary waveguide mode which can be probed with the quantum dots by choosing a sample with $a=260$ nm  and $r=78$ nm, and in this case we focus on $\sigma=0\%.$ We observe a distribution of different Purcell factors reflecting the statistical distribution of coupling coefficients due to the random nature of the Anderson-localized cavities together with uncontrolled spatial and spectral matching of the quantum dot emitters to the cavities. We observe an average Purcell factor of 1.7 together with a variance of 0.5. We stress that the Purcell factor obtained from these types of measurements constitute lower bounds due to the lack of spectral tuning. The histograms in Figs. \ref{3}(b-d) include the experimentally extracted and theoretically expected Purcell factor distributions for the waveguides with parameters $a=240$ nm , $r=74$ nm and $\sigma=0\%$, $1\%$ and $2\%$ respectively. In this case the fundamental waveguide mode is probed, which is expected to lead to the best performance. The observed Purcell factors in this case range from 0.1 to 4.5, i.e., very pronounced suppression and enhancement is observed reflecting the broad range of coupling efficiencies found due to the statistical properties of the cavities. An average Purcell factor ranging between 2.0 and 2.5 is observed depending on the degree of disorder. The localized modes are found to span a spectral range between 3 nm and 7 nm \cite{david_intrinsicdisorder}. Compared to the measurements made at the higher lying waveguide mode, cf. Fig. \ref{3}(a), the Purcell factors are generally found to be higher and have a broader distribution for the fundamental mode where also higher cavity Q-factors are observed, see insets of Figs \ref{3}(a)-(d). Figures \ref{3}(b)-(d) also plot the calculated Purcell factor distributions based on the theory put forward in Ref. \cite{henri_prl}. {In these calculations, we have used $\xi=7$ $\mu${m}, $12$ $\mu${m} and $10$ $\mu${m}  with corresponding loss lengths $l=500$ $\mu${m}, $400$  $\mu${m} and $400$ $\mu${m}, extracted using the method described in \cite{njp_smolka}}. As observed, a very wide Purcell factor distribution is expected from theory, and while the most abundant Purcell factor observed in the experiment matches theory rather well, the complete distribution is not revealed experimentally. A number of reasons could be responsible for these discrepancies apart from the limited statistics of the measurements.{ Thus}, the observation of the high Purcell factors are limited by the lack of control over the spectral detuning in the present experiment, since the theory assumes that the quantum dots are resonant with the cavity modes while being spatially distributed. Furthermore, the applied theory is based on a simple 1D plane-wave  model with experimentally realistic localization length and loss length \cite{henri_prl} while the exact details of the Bloch mode in the PhCW are not accounted, which may give rise to complex nonuniversal light-matter coupling strengths \cite{david_c0}. The low-value tail of the Purcell factor distributions are difficult to extract experimentally primarily due to the fact that quantum dots have two orthogonal bright exciton states and preferentially  emission is collected from the component that couples best to the cavity meaning that the very slow quantum dots are not observed. Furthermore, the cavities with very low Q-factors are not clearly identified experimentally relative to the background emission from the sample. Figure \ref{3}(e) compares calculated and measured mean value of the Purcell factor that are in reasonable agreement despite the complexities discussed above. For completeness also the variance of the Purcell factor, which is a characteristics of the distributions, is plotted in Fig.\ref{3}(e).

\begin{figure}[ht!]
  \includegraphics[width=157mm, height=140mm]{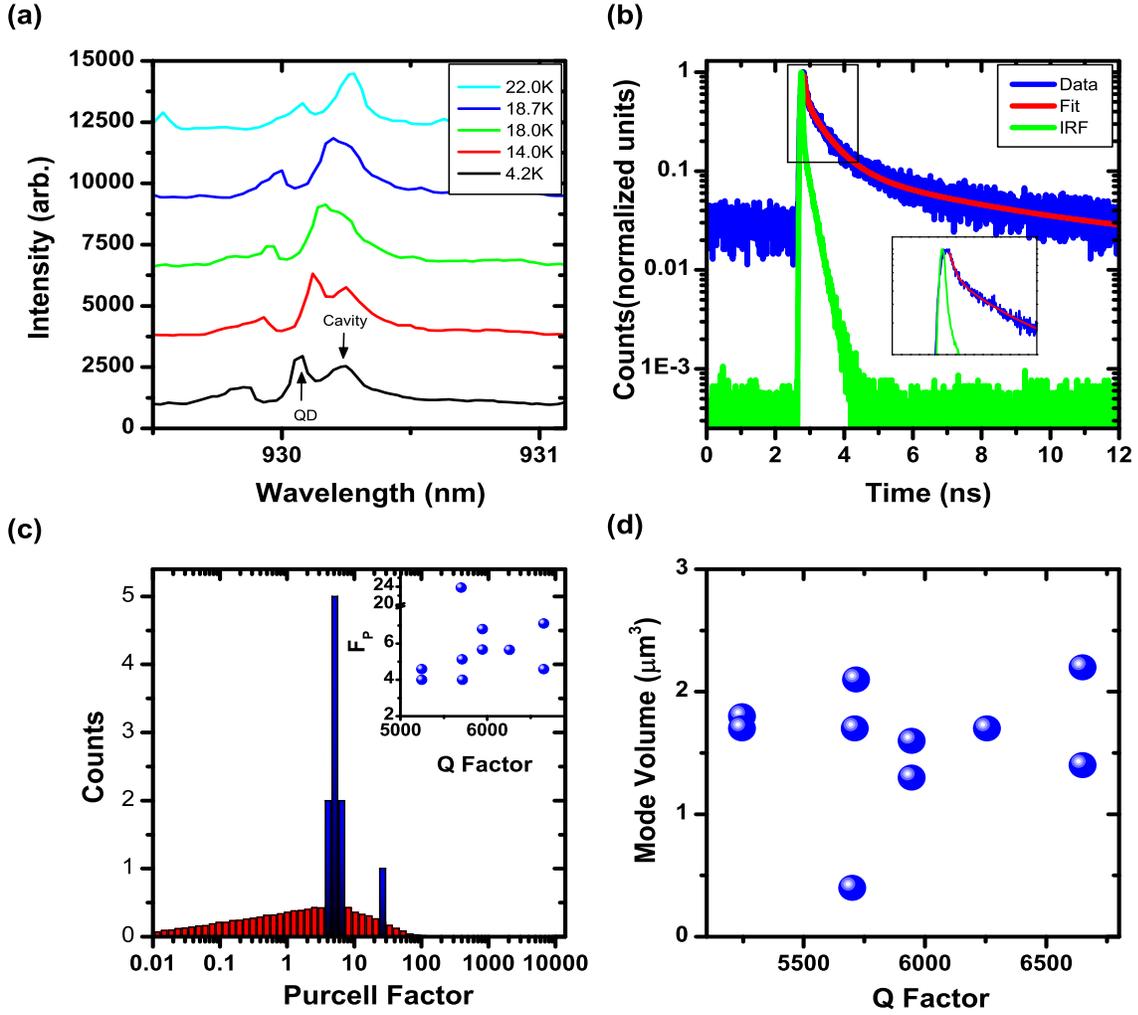}
      \caption{ \label{4} \textbf{Statistical distribution of Purcell factors from single quantum dot on resonance with Anderson-localized cavities.} (a) Emission spectrum of the fastest quantum dot (Purcell factor of 23.8) while temperature tuning it through resonance of an Anderson-localized mode. {(b) Decay curve recorded from the quantum dot in (a) at resonance with the cavity. The fit is shown as the solid red line. The green curve is the instrument response function(IRF) of the detector).}  (c) Purcell factor statistics obtained after tuning single quantum dots into resonance for a PhCW with r = 69 nm, a = 230 nm and $\sigma=1\%$ (Blue histogram). The red histogram shows the theoretically calculated distribution. (d) Extracted upper bound on the mode volume vs. the corresponding cavity Q-factor.}
    \end{figure}

\section{Time-resolved measurements on single quantum dots}

In the previous section, we found a maximum in the average Purcell factor for quantum dots coupled to Anderson-localized cavities appearing near the cutoff of the fundamental guided mode. In the more precise Purcell factor measurements, the detuning between quantum dot and cavity is controlled through temperature and the decay rate is extracted on resonance. Figure \ref{4}(a) shows the spectrum of a single quantum dot while temperature tuning it  across the resonance of an Anderson-localized cavity. We have repeated this procedure for in total 10 different quantum dots along the PhCW. The statistics is plotted in Fig. \ref{4}(c) where we observe Purcell factors in the range of 4 - 7 together with a quantum dot with a Purcell factor as high as 23.8. For comparison, the theoretically predicted distributions are also plotted in Fig. \ref{4}(c), however a quantitative comparison between theory and experiment is limited by the effects discussed in the previous section and is outside the scope of the present work. The inset in Fig. \ref{4}(c) plots the measured Purcell factors vs. the cavity Q, and no clear correlation is observed, which is attributed to the fact that the quantum dots are positioned randomly relative to the electric field of the localized cavities. Using the theoretical expression for the Purcell factor, we can estimate an upper bound on the mode volume of the individual Anderson-localized modes that are plotted in in Fig. \ref{4}(d). The extracted values range between 1.2 to 2.2 $\mu$m\textsuperscript{3}, where we stress that the evidence for significant spatial mismatch from the lack of correlation between Q and the Purcell factor implies that the actual mode volumes are likely to be significantly smaller. Consequently the extracted values are likely to be consistent with the mode volumes recently extracted from random lasing experiments in disordered PhCW, where mode volumes in the range of 0.07 - 0.1  $\mu$m\textsuperscript{3} were derived \cite{jin}.

Finally we analyze in detail the case of a Purcell factor of 23.8. In this case the upper bound for the mode volume is 0.4 ${\mu}m^3$ and the spatial positioning and dipole orientation is likely to be close to optimal. The criterion for strong coupling between a quantum dot and a cavity is $g/\kappa>1/4$ \cite{henri_prl}, {where $\kappa=2{\pi}c/{\lambda}Q$ is the loss rate of the cavity, $g\approx \sqrt{\Gamma/4 \kappa}$ is the coupling strength between the emitter and the cavity, $\Gamma$ is the decay rate of the emitter in the weak-coupling regime, $c$ is speed of light in vacuum, and $\lambda$ is the wavelength of emitted photon}. For this particularly fast quantum dot this ratio is $g/\kappa=0.13$, which indicates that the cavity is in the weak-coupling regime, but close to the onset of strong coupling.

\section{Conclusions}

In conclusion, we have presented a statistical analysis of the emission dynamics from single quantum dots embedded in disordered PhCWs. The measurement of the decay dynamics from Anderson-localized modes enables us to efficiently collect a large amount of data. Such a statistical analysis can be used to identify the structural parameters and degree of disorder that gives rise to the most pronounced cavity QED effects. Measuring the decay rate of single quantum dots that are spectrally tuned across the cavities offer reliable extraction of the Purcell factor. We observed Purcell factors in the range of 4-7 together with an extraordinary large Purcell factor of 23.8. A very broad distribution of Purcell factors is predicted from theory and the present work constitutes the first experimental report on systematic studies of the Purcell factor statistics in Anderson-localized cavities.

\section{Acknowledgments}

We gratefully acknowledge financial support from the Danish council for independent research (natural sciences and technology and production sciences), and the European research council (ERC consolidator grant "ALLQUANTUM").

\end{document}